\documentclass[prd,onecolumn,nofootinbib,aps,tightenlines,preprintnumbers,notitlepage,longbibliography,superscriptaddress,12pt]{revtex4-1}

\usepackage{multirow}
\usepackage{hyperref}
\usepackage{graphicx}
\usepackage{bm}
\usepackage{color, colortbl, xcolor}

\newcommand{\be}{\begin{eqnarray}}
\newcommand{\ee}{\end{eqnarray}}
\newcommand{\bn}{\bm{n}}

\newcommand{\unlensed}{\mathrm{u}}

\newcommand{\caps}{Center for AstroPhysical Surveys, National Center for Supercomputing Applications, Urbana, IL, 61801, USA}

\begin{document}

\title{The Impact of Cross-Covariances Between the CMB and Reconstructed Lensing Power}

\author{Cynthia~Trendafilova}
\affiliation{\caps}

\date{\today}

\begin{abstract}
Weak gravitational lensing of the Cosmic Microwave Background (CMB) changes CMB statistics in a nontrivial way, allowing for reconstruction of the lensing potential and the use of these reconstructed maps in determining cosmological parameters that affect the formation of intervening large-scale structures. Although in principle there are correlations between the primary CMB and the reconstructed lensing potential due to the lensing procedure itself, in practice CMB analyses treat these as negligible when combining these band powers in likelihoods. In this paper we quantify explicitly the impact on parameter constraints due to these cross-covariances between the lensed CMB and reconstructed lensing power, and we compare to the effect of including all lensing-induced non-Gaussian covariances, which have previously shown to impact parameter constraints on the order of 10\%. We perform our analysis for a range of experimental setups, scanning over instrumental noise levels of 0.5 to 10.0 $\mu$K-arcmin in temperature assuming fully polarized detectors, and using a fixed beam size of 1.4 arcmin. When the correlations between the lensed CMB and lensing power are neglected, we find that forecasted constraints shift by at most 3\% of the error bar for a 6-parameter $\Lambda$CDM model, and for the noise levels considered in this paper. For some of the $\Lambda$CDM extensions considered here, however, these correlations have a nontrivial impact, in some cases more than 10\% of the error bar, even for current experimental noise levels.
\end{abstract}

\maketitle

\section{Introduction}
\label{sec:Introduction}

Observations of the Cosmic Microwave Background (CMB) have become an increasingly powerful tool for learning about the history and properties of our universe. Current experiments have provided tight constraints on cosmological parameters, and upcoming data sets from surveys such as Simons Observatory and CMB-S4 will provide even higher precision measurements of the CMB.

Although we might hope for an unobstructed view of the CMB in order to most accurately measure its anisotropies, CMB photons traveling to us from the last-scattering surface interact with the matter and structures that have formed in our universe. One significant effect is weak gravitational lensing, causing photons to have their paths deflected by large-scale structures. This affects the primary CMB in well-understood ways: it smooths the peaks of the power spectra, transfers power from large to small angular scales according to the size of the features causing the lensing, and converts $E$-mode power into $B$-mode power~\cite{Lewis:2006fu}. Although on the one hand this distorts our view of the last scattering surface and creates additional difficulty in accurately determining the CMB anisotropies prior to the photons' deflection, it also encodes information regarding the intervening structures that have caused the lensing. The statistics of the CMB are altered in a way that allows us to reconstruct the lensing potential responsible for the deflections~\cite{Hu:2001kj, Okamoto:2003zw}. This weak gravitational lensing effect has been detected with surveys such as Planck~\cite{Planck:2018lbu}, ACT~\cite{ACT:2023dou}, and SPT~\cite{Wu:2019hek}, and it will become even more important at the low noise and high resolution of upcoming experiments such as Simons Observatory~\cite{SimonsObservatory:2018koc}, CMB-S4~\cite{CMB-S4:2016ple}, PICO~\cite{NASAPICO:2019thw}, and CMB-HD~\cite{Sehgal:2019ewc}.

By changing the statistics of the primary CMB, weak gravitational lensing results in additional correlations which would not be present in the absence of lensing~\cite{Smith:2004up, Smith:2005ue, Smith:2006nk, Li:2006pu, 2012PhRvD..86l3008B, Schmittfull:2013uea}. Furthermore, because the lensing reconstruction procedure utilizes the lensed CMB itself, this inevitably leads to correlations between the measured lensed CMB and the reconstructed lensing power. Lensing is therefore responsible for both coupling $\ell$ modes across different angular scales, and for correlating the lensing power spectrum with the lensed primary CMB power spectra. The effect of these lensing-induced covariances, and whether or not they are accounted for in determining parameters, has shown to impact error bars by several tens of percent for an experiment like CMB-S4~\cite{Peloton:2016kbw}. One can delens the primary CMB to mitigate these effects, and getting improved parameter constraints via delensing will become more important for future experiments reaching lower noise levels. Delensing procedures are not perfect and some residual non-Gaussian correlations can remain due to filtering, but their impact on parameter constraints is negligible~\cite{Green:2021xzn}.

Although we can learn about the $\Lambda$CDM model from the anisotropies in the primary CMB, we get additional information, for parameters such as $A_\mathrm{s}$ and $\tau$, by also including the results of our lensing reconstruction. There are further parameters beyond $\Lambda$CDM that also benefit particularly from accurately reconstructing the lensing potential, such as measurements of the neutrino mass~\cite{Dolgov:2002wy, Kaplinghat:2003bh, Lesgourgues:2006nd, Dvorkin:2019jgs, Green:2021xzn} and determination of the dark energy equation of state~\cite{Frieman:2008sn}. In performing analyses with CMB data, one can combine the measurements of the CMB with those of the reconstructed lensing potential and use the combined likelihood to inform the determination of cosmological parameters. In such analyses, it is important to know to what extent these two sets of information can be combined as if they are independent of one another, and how correlations between the lensing power and the lensed CMB power need to be modeled and accounted for.

Because lensing-induced non-Gaussian covariances are coupling CMB modes that were otherwise assumed to be independent of one another, they reduce the information and constraining power of CMB data. If one neglects them entirely and assumes only Gaussian covariances among CMB power spectra, resulting parameter constraints will be overly optimistic; previous works~\cite{Peloton:2016kbw, Green:2021xzn} have shown that neglecting the non-Gaussian covariances can underestimate constraints by 10\% or more for some parameters. The authors in~\cite{Motloch:2016zsl} studied the importance of including the covariances between the lensed CMB and the reconstructed lensing potential for an experimental setup like that of CMB-S4, and they found that these terms impact forecasted parameter constraints by no more than 4\% for 6-parameter $\Lambda$CDM, but up to 20\% for some additional parameter extensions.

In our work, we forecast parameter constraints over a range of noise levels from 0.5 to 10.0 $\mu$K-arcmin in temperature, in order to study the relevance of these effects for current CMB-S4 surveys as well as future experiments. We also include additional terms in the non-Gaussian covariance, proportional to the derivatives of the lensed CMB power spectra with respect to the unlensed CMB, for all modes, including $TT$, $TE$, and $EE$. This term is particularly relevant in the case of the $BB$ power spectrum, since this power is generated entirely by the lensing of $E$-modes in the absence of primordial $B$-modes. However, the terms dependent on the other power spectra also have a non-trivial impact on the covariance matrix and the resulting parameter constraints, especially for parameters whose constraints are driven primarily by measurements of the lensing power~\cite{Hotinli:2021umk}. We therefore use a covariance model which includes these terms to quantify how much of the overall impact of non-Gaussian lensing-induced covariances is due specifically to the cross-covariances between the primary CMB power spectra and the power spectrum of the reconstructed lensing potential. We find that this effect, for 6-parameter $\Lambda$CDM, contributes no more than 3\% out of a total order 10\% effect, such that the off-diagonal $\ell \times \ell$ correlations amongst primary CMB modes provide the more impactful contribution in the full covariance matrix. For $\Lambda$CDM extensions whose parameters are largely constrained by lensing information, we find that these correlations can be relevant at the order of 10\% or more even at noise levels of 10.0 $\mu$K-arcmin.

We perform our calculations using the publicly available codes \texttt{CLASS\_delens}\footnote{\url{https://github.com/selimhotinli/class_delens}}, which calculates lensed CMB power spectra, and then finds delensed CMB power spectra and lensing reconstruction noise using an iterative quadratic estimator, and \texttt{FisherLens}\footnote{\url{https://github.com/ctrendafilova/FisherLens}}, which allows for Fisher forecasting using CMB covariance matrices that include lensing-induced non-Gaussian covariances~\cite{Hotinli:2021umk}.

Our paper is structured as follows. In Sec.~\ref{sec:Covariance}, we present the model used to implement lensing-induced non-Gaussian covariances in \texttt{FisherLens}, while in Sec.~\ref{sec:Forecasts} we outline the details of our forecasting procedure. We present our results in Sec.~\ref{sec:Results}, and we conclude in Sec.~\ref{sec:Conclusion}.

\section{Lensing-Induced Covariance in the CMB}
\label{sec:Covariance}
The cosmic microwave background power spectra, $C_\ell$, quantify the magnitude of the fluctuations in the CMB power across different angular scales, $\ell$. In the absence of primordial non-Gaussianities in the fluctuations of the early universe, we expect the statistics of the primary CMB anisotropies to be well described as Gaussian. In this case, the fluctuations at a particular angular scale, $\ell_1$, are not correlated with the fluctuations at a different angular scale, $\ell_2$. If one considers a two-dimensional covariance matrix describing the correlations amongst all possible modes, then over the full-sky the entries are diagonal, such that covariance between temperature and/or polarization fluctuations at a given $\ell$ depends only on our power spectra and the noise, $N_\ell$, with which we have measured them.

Gravitational lensing of CMB photons distorts the features in our CMB maps, shifting power from large angular scales to small ones, and thereby generating non-zero connected four-point functions which we must model in addition to the two-point statistics~\cite{Zaldarriaga:2000ud, Hu:2001fa}. The CMB power across different angular scales becomes correlated, generating off-diagonal elements in the covariance matrix. Additionally, because we reconstruct lensing maps from our measurements of the CMB itself, the reconstruction procedure results in correlations between the lensing power and the power spectrum of the primary CMB.

To properly model these covariances, we must consider how varying the underlying physics affects the lensed CMB. Consider a photon that we observe from a line-of-sight direction $\bn$ after it has been deflected by some angle $\mathbf{d}(\bn)$, where the deflection angle is given in terms of the lensing potential $\phi$ as $\mathbf{d}(\bn) = \mathbf{\nabla}\phi(\bn)$. Then to lowest order, the observed lensed CMB temperature map, $\tilde{T}$, depends on the unlensed map $T$ according to
\begin{equation}
    \tilde{T}(\bn)=T(\bn+\mathbf{d}(\bn))\simeq T(\bn) + \mathbf{d}(\bn)\cdot\mathbf{\nabla}T(\bn) + \ldots \, .
    \label{eq:T_lens}
\end{equation}
Clearly for a given lensing potential, the lensed temperature map will change if the unlensed map changes. Furthermore, if we change the lensing potential while keeping the unlensed CMB fixed, the lensed CMB changes as well. These derivatives contribute to the power spectrum covariance matrix between different $\ell$ modes, which is implemented in \texttt{FisherLens} according to the model of~\cite{Benoit-Levy:2012dqi}
\be
    {\rm Cov}_{\ell_1\ell_2}^{XY,WZ}=&&  f_{\rm sky}^{-1} \Bigg\{ \frac{\delta_{\ell_1\ell_2}}{2\ell_1+1} \left[ \left( C_{\ell_1}^{XY} + N_{\ell_1}^{XY} \right) \left( C_{\ell_1}^{WZ} + N_{\ell_1}^{WZ} \right) + 
    \left( C_{\ell_1}^{XW} + N_{\ell_1}^{XW} \right) \left( C_{\ell_1}^{YZ} + N_{\ell_1}^{YZ} \right)
    \right] \nonumber \\
    &&+\sum\limits_\ell \left(\frac{\partial C_{\ell_1}^{XY}}{\partial {C}_{\ell}^{XY,\unlensed}} \widehat{{\rm Cov}}_{\ell\ell}^{X^{\unlensed} Y^{\unlensed} , W^{\unlensed} Z^{\unlensed}} \frac{\partial C_{\ell_2}^{WZ}} {\partial C_{\ell}^{WZ, \unlensed}} \right) (1-\delta_{\ell_1 \ell_2})
    \nonumber\\
    &&+\sum\limits_\ell \left(\frac{\partial C_{\ell_1}^{XY}}{\partial {C}_{\ell}^{\phi\phi}} \widehat{{\rm Cov}}_{\ell\ell}^{\phi\phi,\phi\phi}\frac{\partial C_{\ell_2}^{WZ}}{\partial {C}_{\ell}^{\phi\phi}}\right)
    (1-\delta_{X\phi}\delta_{Y\phi}\delta_{Z\phi}\delta_{W\phi} )
    \Bigg\} \, .
    \label{eq:covarianceNG}
\ee
Here we let $X\in\{T,E,B,\phi\}$ and $X^\unlensed\in\{T^\unlensed,E^\unlensed,B^\unlensed\}$ and similarly for $Y$, $W$, and $Z$. The contributions to the covariance due to the non-Gaussian correlations are given by the second and third lines. There is an additional term linear in $C_\ell^{\phi\phi}$ which we have neglected here, as it is numerically small~\cite{Peloton:2016kbw}. $\widehat{{\rm Cov}}_{\ell\ell}^{XY,WZ}$ refers to the noiseless Gaussian power spectrum covariance, given by 
\be
    \widehat{{\rm Cov}}_{\ell_1\ell_2}^{XY,WZ}=&&  \frac{\delta_{\ell_1\ell_2}}{2\ell_1+1} \left[ C_{\ell_1}^{XY} C_{\ell_1}^{WZ} + 
    C_{\ell_1}^{XW} C_{\ell_1}^{YZ} \right] .
    \label{eq:covarianceGNoiseless}
\ee
The terms in the second line of eq.~\ref{eq:covarianceNG} include the contribution from $B$-mode power that has been generated by lensing of $E$-modes. Additionally, these terms have a non-trivial impact on covariances between $TT$, $TE$, and $EE$ power spectra, and they have been shown to be important for getting accurate parameter constraints, especially in the case of parameters such as $A_s$ and $\tau$~\cite{Hotinli:2021umk}. The full checkerboard pattern of correlations amongst different $\ell$ modes of the primary CMB, along with the portion coming from the contribution of the second line in eq.~\ref{eq:covarianceNG}, can be seen in fig.~6 of \cite{Hotinli:2021umk}.

Eq.~\ref{eq:covarianceNG} is meant to be a compact way of writing the covariance for all combinations of $X\in\{T,E,B,\phi\}$. For clarity, we break this down explicitly for different combinations of the primary CMB, $I,J,W,V\in\{T,E,B\}$, and $\phi$. For the correlations amongst $\ell$-modes of the CMB, we have
\be
    {\rm Cov}_{\ell_1\ell_2}^{IJ,WV}=&&  f_{\rm sky}^{-1} \Bigg\{ \frac{\delta_{\ell_1\ell_2}}{2\ell_1+1} \left[ \left( C_{\ell_1}^{IJ} + N_{\ell_1}^{IJ} \right) \left( C_{\ell_1}^{WV} + N_{\ell_1}^{WV} \right) + 
    \left( C_{\ell_1}^{IW} + N_{\ell_1}^{IW} \right) \left( C_{\ell_1}^{JV} + N_{\ell_1}^{JV} \right)
    \right] \nonumber \\
    &&+\sum\limits_\ell \left(\frac{\partial C_{\ell_1}^{IJ}}{\partial {C}_{\ell}^{IJ,\unlensed}} \widehat{{\rm Cov}}_{\ell\ell}^{I^{\unlensed} J^{\unlensed} , W^{\unlensed} V^{\unlensed}} \frac{\partial C_{\ell_2}^{WV}} {\partial C_{\ell}^{WV, \unlensed}} \right) (1-\delta_{\ell_1 \ell_2})
    \nonumber\\
    &&+\sum\limits_\ell \left(\frac{\partial C_{\ell_1}^{IJ}}{\partial {C}_{\ell}^{\phi\phi}} \widehat{{\rm Cov}}_{\ell\ell}^{\phi\phi,\phi\phi}\frac{\partial C_{\ell_2}^{WV}}{\partial {C}_{\ell}^{\phi\phi}}\right)
    \Bigg\} \, ,
    \label{eq:covarianceXXXX}
\ee
while for the correlations between the CMB and the lensing power spectrum, we have
\be
    {\rm Cov}_{\ell_1\ell_2}^{IJ,\phi\phi}=&&  f_{\rm sky}^{-1} \Bigg\{ \frac{\delta_{\ell_1\ell_2}}{2\ell_1+1} \left[ \left( C_{\ell_1}^{IJ} + N_{\ell_1}^{IJ} \right) \left( C_{\ell_1}^{\phi\phi} + N_{\ell_1}^{\phi\phi} \right) 
    \right] \nonumber \\
    &&+\sum\limits_\ell \left(\frac{\partial C_{\ell_1}^{IJ}}{\partial {C}_{\ell}^{\phi\phi}} \widehat{{\rm Cov}}_{\ell\ell}^{\phi\phi,\phi\phi}\frac{\partial C_{\ell_2}^{\phi\phi}}{\partial {C}_{\ell}^{\phi\phi}}\right)
    \Bigg\} \, .
    \label{eq:covarianceXXdd}
\ee
Finally, the Gaussian covariance for different $\ell$ modes of $\phi$ is given by
\be
    {\rm Cov}_{\ell_1\ell_2}^{\phi\phi,\phi\phi}=&&  f_{\rm sky}^{-1}  \frac{\delta_{\ell_1\ell_2}}{2\ell_1+1} \left[ \left( C_{\ell_1}^{\phi\phi} + N_{\ell_1}^{\phi\phi} \right) \left( C_{\ell_1}^{\phi\phi} + N_{\ell_1}^{\phi\phi} \right) + 
    \left( C_{\ell_1}^{\phi\phi} + N_{\ell_1}^{\phi\phi} \right) \left( C_{\ell_1}^{\phi\phi} + N_{\ell_1}^{\phi\phi} \right)
    \right]
     \, .
    \label{eq:covariancedddd}
\ee
To facilitate visualization of the covariance matrix, we define the correlation matrix
\begin{equation}
    R_{\ell_1 \ell_2}^{XY,WZ} \equiv \frac{ \mathrm{Cov}_{\ell_1\ell_2}^{XY,WZ} } {\sqrt{\mathrm{Cov}_{\ell_1\ell_1}^{XY,XY} \mathrm{Cov}_{\ell_2\ell_2}^{WZ,WZ}} } \, ,
    \label{eq:correlationMatrix}
\end{equation}
where each element is normalized according to the corresponding diagonal elements, allowing us to easily gauge the relative size of the off-diagonal terms.

\begin{figure}[t!]
    \centering
    \vspace{-0.45cm}
    \includegraphics[width=1.00\columnwidth]{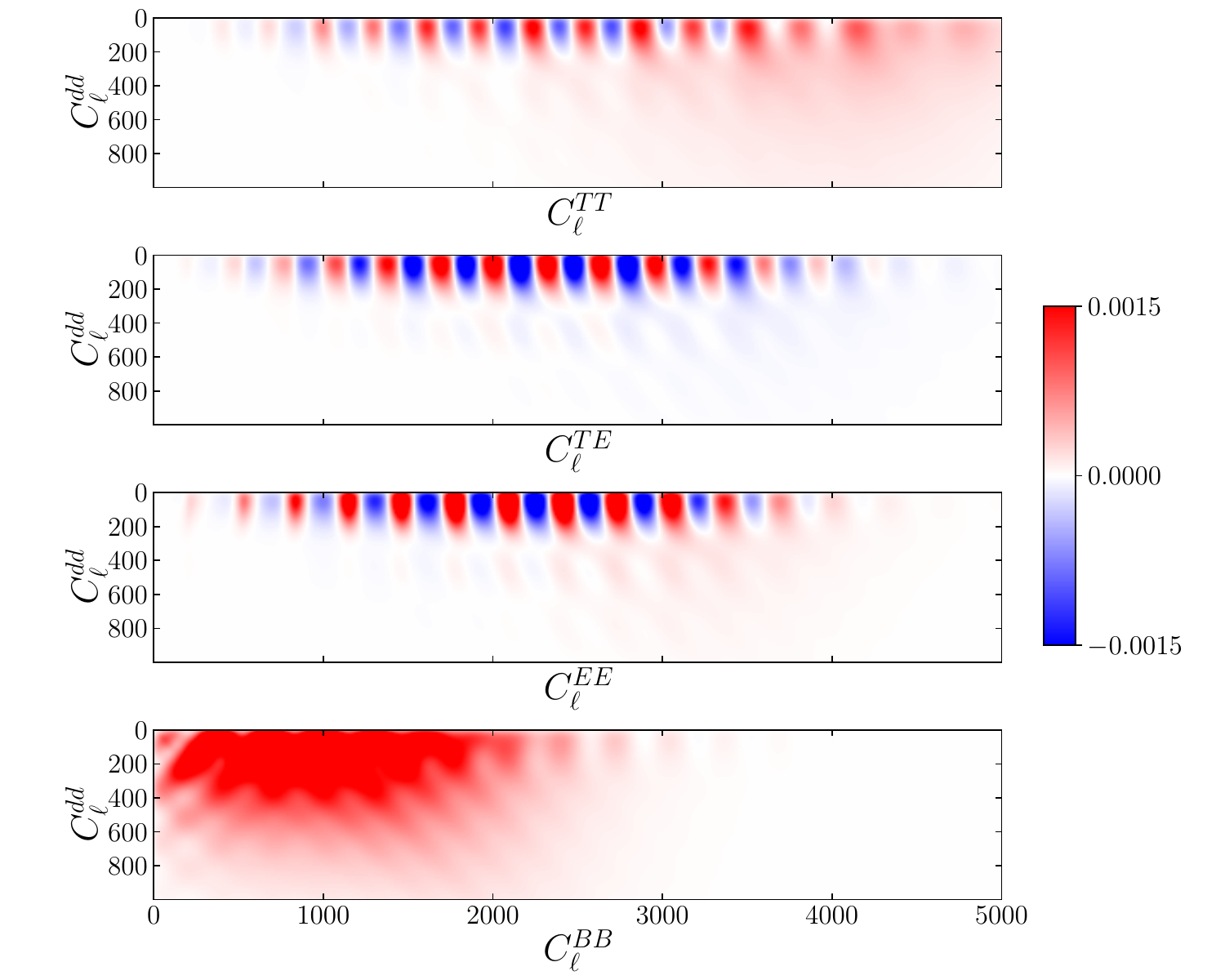}
    \vspace{-0.45cm}
    \caption{Covariance of eq.~\ref{eq:covarianceXXdd} scaled according to eq.~\ref{eq:correlationMatrix}, for an experiment of 1 $\mu K$-arcmin white noise in the temperature and a 1.4 arcmin beam width. Note that the scale is saturated at +0.001 and -0.001 to facilitate visibility of the correlation pattern; the actual maximum and minimum values are within +/- 0.005. The complex correlation structure results in both positive and negative values for the correlation of $C_\ell^{dd}$ with all power spectra except $C_\ell^{BB}$, which is generated entirely by the conversion of $E$-modes via lensing.}
    \label{fig:Correlations}
\end{figure}

Although this can of course be plotted for the entire correlation matrix, we focus in figure~\ref{fig:Correlations} on the correlations of eq.~\ref{eq:covarianceXXdd}, in particular, since we would like to evaluate their relevance for constraining parameters from CMB data. All power spectra and derivatives are calculated with \texttt{CLASS\_delens}; lensing reconstruction noise is calculated by iteratively delensing all CMB spectra and calculating the final minimum-variance reconstruction noise using all quadratic estimators ($TT$, $TE$, $TB$, $EE$, and $EB$). The resulting correlations between $C_\ell^{BB}$ and $C_\ell^{dd}$ are different from the other modes since the $B$-modes are generated entirely by the lensing. For all panels, we plot the correlation values for multipoles up to 1000 of the reconstructed lensing potential, where the correlations are most prominent. The correlations are largest at $\ell$ of a few hundred in $dd$, but are nonetheless numerically small, less than 1\%, when compared to the diagonal Gaussian covariances of the $dd$ and primary CMB power spectra. However, we do not assume that their numerically small values necessarily mean that they are negligible for constraining parameters, especially for a range of experimental noise levels and cosmological models. The full non-Gaussian covariance contributions in eq.~\ref{eq:covarianceNG} result in correlations that are likewise numerically small but have been shown to impact parameter constraints on the order of 10\%. Therefore, in order to better quantify the relative impact of the terms in eq.~\ref{eq:covarianceXXdd}, we perform Fisher forecasts as detailed in the next section, and evaluate their effect at the level of parameter error bars.

\section{Parameter Forecasts}
\label{sec:Forecasts}

To quantify the importance of the correlations between the CMB and the lensing power in eq.~\ref{eq:covarianceXXdd}, we use Fisher forecasting to get parameter constraints, both with and without these correlations included. We model the experiment's temperature noise spectrum as white noise, according to
\begin{equation}
    N_{\ell}^{TT} = \Delta_T^2 \, \mathrm{exp} \left( \ell(\ell+1) \frac{\theta_{\mathrm{FWHM}}^2}{8\log{2}} \right) \, ,
    \label{eq:TT_noise}
\end{equation}
where $\Delta_T$ is the instrumental noise in $\mu$K-rad, $\theta_{\mathrm{FWHM}}$ is the full-width at half-maximum beam size in radians, and we assume the detectors are fully polarized such that $N_{\ell}^{EE} = N_{\ell}^{BB} = 2N_{\ell}^{TT}$; we do not include foregrounds. To capture the relevance of these terms for both current and upcoming CMB surveys, we scan over a range of noise levels $\Delta_T$ from 0.5 to 10.0 $\mu$K-arcmin, with a beam size $\theta_{\rm FWHM}$ fixed at 1.4 arcmin and a sky fraction of 0.5.

\begin{table}
\begin{center}
 \begin{tabular}{l@{\hskip 12pt}l @{\hskip 12pt}c@{\hskip 12pt}c} 
 \toprule
   Parameter & Symbol    &   Fiducial Value      & Step Size     \\ [0.5ex] 
 \hline
Physical cold dark matter density &   $\Omega_\mathrm{c} h^2$ &   0.1197 	            & 0.0030 	    \\ 
Physical baryon density &   $\Omega_\mathrm{b} h^2$ &   0.0222 	            & $8.0\times10^{-4}$ 	    \\
Angle subtended by acoustic scale &   $\theta_\mathrm{s}$     &   0.010409 	            & $5.0\times10^{-5}$ 	    \\
Thomson optical depth to recombination &   $\tau$         &   0.060 	            & 0.020 	    \\
Primordial scalar fluctuation amplitude &   $A_\mathrm{s}$          &   $2.196\times10^{-9}$  & $0.1\times10^{-9}$ 	    \\
Primordial scalar fluctuation slope &   $n_\mathrm{s}$          &   0.9655 	            & 0.010 	    \\
   \hline
Tensor-to-scalar ratio & $r$ & 0.01 & 0.001\\
Primordial tensor fluctuation tilt & $n_\mathrm{t}$ & -0.00129 & 0.00129 \\
\hline
Effective number of neutrino species & $N_\mathrm{eff}$ & 3.046 & 0.080 \\
Sum of neutrino masses & $\Sigma m_{\nu}$ (meV) & 60 & 20 \\
Curvature density & $\Omega_\mathrm{k}$ & 0 & 0.01 \\
Dark energy equation of state & $w_0$ & -1.0 & 0.3 \\
  \hline
\end{tabular}
    \caption{
    Fiducial cosmological parameters and step sizes for numerical derivatives used in forecasts, taken from~\cite{Allison:2015qca}. All of our forecasts in Sec.~\ref{sec:Results} include the 6-parameter $\Lambda$CDM model given in the table; we additionally consider several extensions, namely $r$, $n_\mathrm{t}$, $N_\mathrm{eff}$, $\Sigma m_{\nu}$, $\Omega_\mathrm{k}$, and $w_0$, which are specified in the text when used.
    }
\label{table:cosmo_fiducial}
\end{center}
\end{table}

The Fisher matrix including the primary CMB and lensing power is given by
\begin{equation}
    F_{ij} = \sum\limits_{\ell_1, \ell_2} \ \sum\limits_{W X Y Z} 
    \frac{\partial C_{\ell_1}^{XY}}{\partial \lambda^i} 
    \left[ \mathrm{Cov}_{\ell_1\ell_2}^{XY,WZ} \right]^{-1}
    \frac{\partial C_{\ell_2}^{WZ}}{\partial \lambda^j} \, ,
    \label{eq:FisherMatrix}
\end{equation}
where $C_\ell^{dd} = \ell(\ell+1)C_\ell^{\phi\phi}$. Unless otherwise specified, we include $TT$, $TE$, $EE$, and $dd$ power spectra. We sum from $\ell = 30$ to 5000 for $TE$ and $EE$, while including only up to $\ell = 3000$ in $TT$ in order to account for the effects of astrophysical foregrounds, since we do not add them in our noise model. For the lensing deflection power spectrum $dd$, we include angular multipoles up to $L = 5000$; we consider three different minimum $L$ values of 2, 50, and 100. Lensing reconstruction is performed including only up to $\ell$ = 3000 in $TT$ in order to mitigate biases from the foregrounds at higher $\ell$. The cosmological parameters considered in our forecasts, along with their fiducial values and step sizes used for taking numerical derivatives, are listed in table~\ref{table:cosmo_fiducial}; we include a $\tau$ prior of $\sigma_\tau = 0.007$. When not being varied, the sum of the neutrino masses is fixed at 60 meV. The primordial helium abundance is set such that it is consistent with standard Big Bang nucleosynthesis predictions.

\section{Results}
\label{sec:Results}

We compare the results of forecasts where the lensing-induced non-Gaussian covariances of our model are included to forecasts where ${\rm Cov}_{\ell_1\ell_2}^{IJ,\phi\phi}$ is neglected. We additionally compare these to the case of neglecting the lensing-induced non-Gaussian covariances in their entirety. For a 6-parameter $\Lambda$CDM model and the range of noise levels considered, we find that using only the Gaussian covariances can underestimate constraints by as much as 10\% for some parameters, such as $\Omega_\mathrm{b} h^2$, consistent with results from previous work~\cite{Peloton:2016kbw, Green:2021xzn}. The effect of including the cross-covariances between the primary CMB and the reconstructed lensing power, however, is small in comparison, shifting parameter constraints by less than 2.5\% of the error bar when ${\rm Cov}_{\ell_1\ell_2}^{IJ,\phi\phi}$ is neglected.

In addition to $\Lambda$CDM, we explore several extended models, with parameters which are informed to varying degrees by the lensing power spectrum. We first consider an 8-parameter model with $r$ and $n_\mathrm{t}$; in this case, we include $B$-modes in the sum of eq.~\ref{eq:FisherMatrix}, from $\ell$ = 30 to 5000. The constraints on $r$ and $n_t$ come primarily from CMB B-modes, and there is no significant degeneracy between these two parameters and $\Lambda$CDM. Thus we expect that including or neglecting the correlations of the lensing power spectrum with the primary CMB will have little effect on parameter constraints for any of the eight parameters in the model; we indeed find that neglecting ${\rm Cov}_{\ell_1\ell_2}^{IJ,\phi\phi}$ shifts the error bars on parameter constraints by at most 3\%.

\begin{figure}[t!]
    \centering
    \includegraphics[width=0.95\columnwidth]{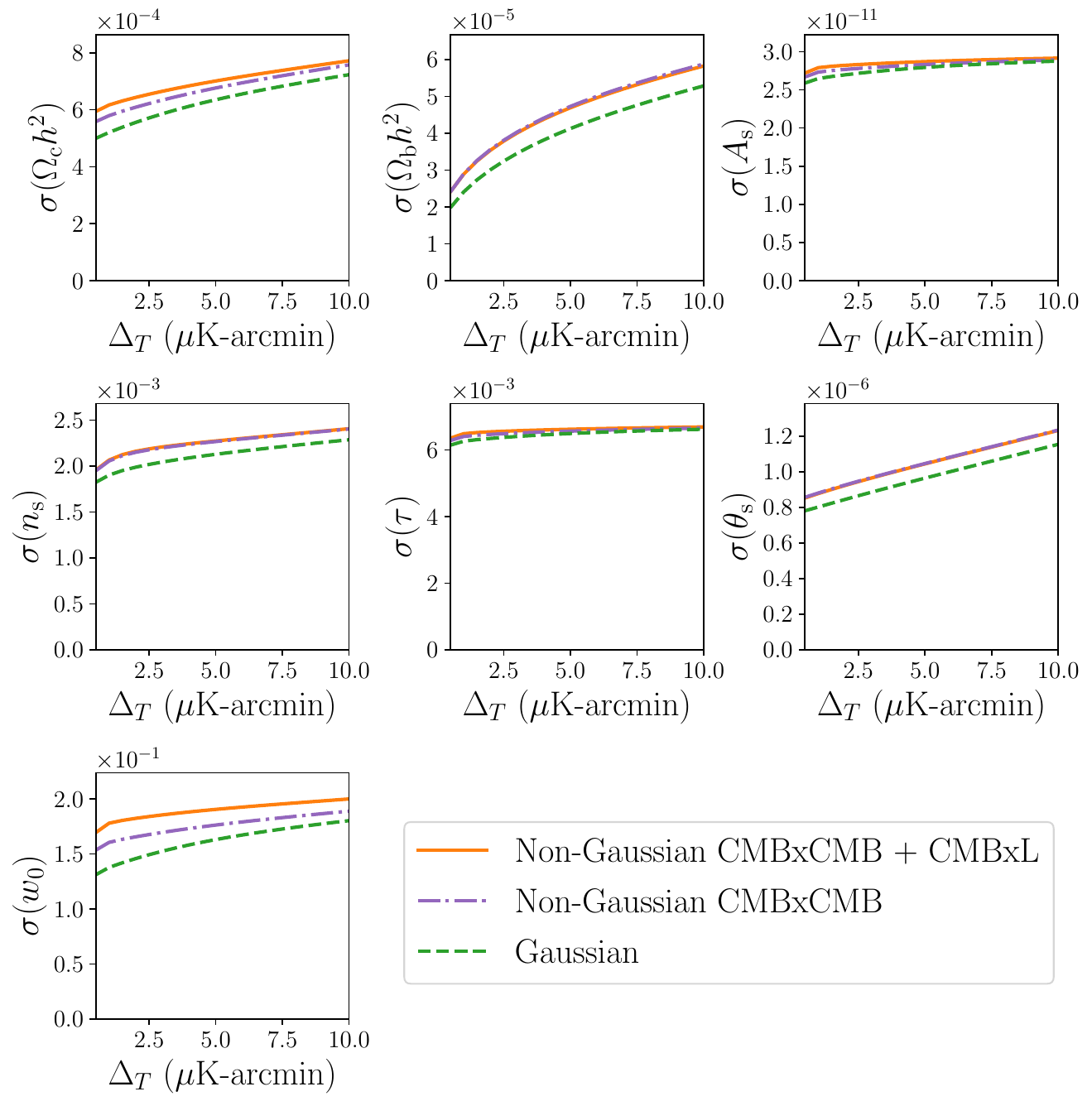}
    \caption{Parameter constraints from lensed CMB spectra as a function of noise level, for a 7-parameter $\Lambda$CDM + $w_0$ model. The dashed green line corresponds to including Gaussian covariances only, the dashed-dotted purple line includes lensing-induced non-Gaussian covariances amongst primary CMB modes, and the solid orange line additionally includes lensing-induced non-Gaussian covariances between the primary CMB and the reconstructed lensing power. The beam size is fixed at 1.4 arcmin for all noise levels. The full non-Gaussian covariance model affects parameter constraints on the order of 10\%, but for most parameters, the covariance of the lensed CMB and lensing power results in a much smaller effect.}
    \label{fig:ParamConstraintsLensed_w}
\end{figure}

We further consider four other parameters to be added to $\Lambda$CDM, which are listed at the bottom of table~\ref{table:cosmo_fiducial}: the effective number of neutrino species $N_\mathrm{eff}$, the sum of the neutrino masses $\Sigma m_{\nu}$, the curvature density $\Omega_\mathrm{k}$, and the dark energy equation of state $w_0$, with the latter three benefitting in particular from lensing information. We find that the addition of $N_\mathrm{eff}$ to our free parameters has little effect on the results, with the relative importance of ${\rm Cov}_{\ell_1\ell_2}^{IJ,\phi\phi}$ being small for $N_\mathrm{eff}$, and relatively unchanged for the other free parameters in the forecast whether $N_\mathrm{eff}$ is included or not. Thus we turn our attention for the rest of this section to the remaining three parameter extensions: $\Sigma m_{\nu}$, $\Omega_\mathrm{k}$, and $w_0$. Different parameter combinations are informed by low-$L$ lensing information in different ways, and the inclusion or exclusion of the correlations of figure~\ref{fig:Correlations} affects parameter constraints to different degrees.

\begin{figure}[t!]
    \centering
    \includegraphics[width=0.95\columnwidth]{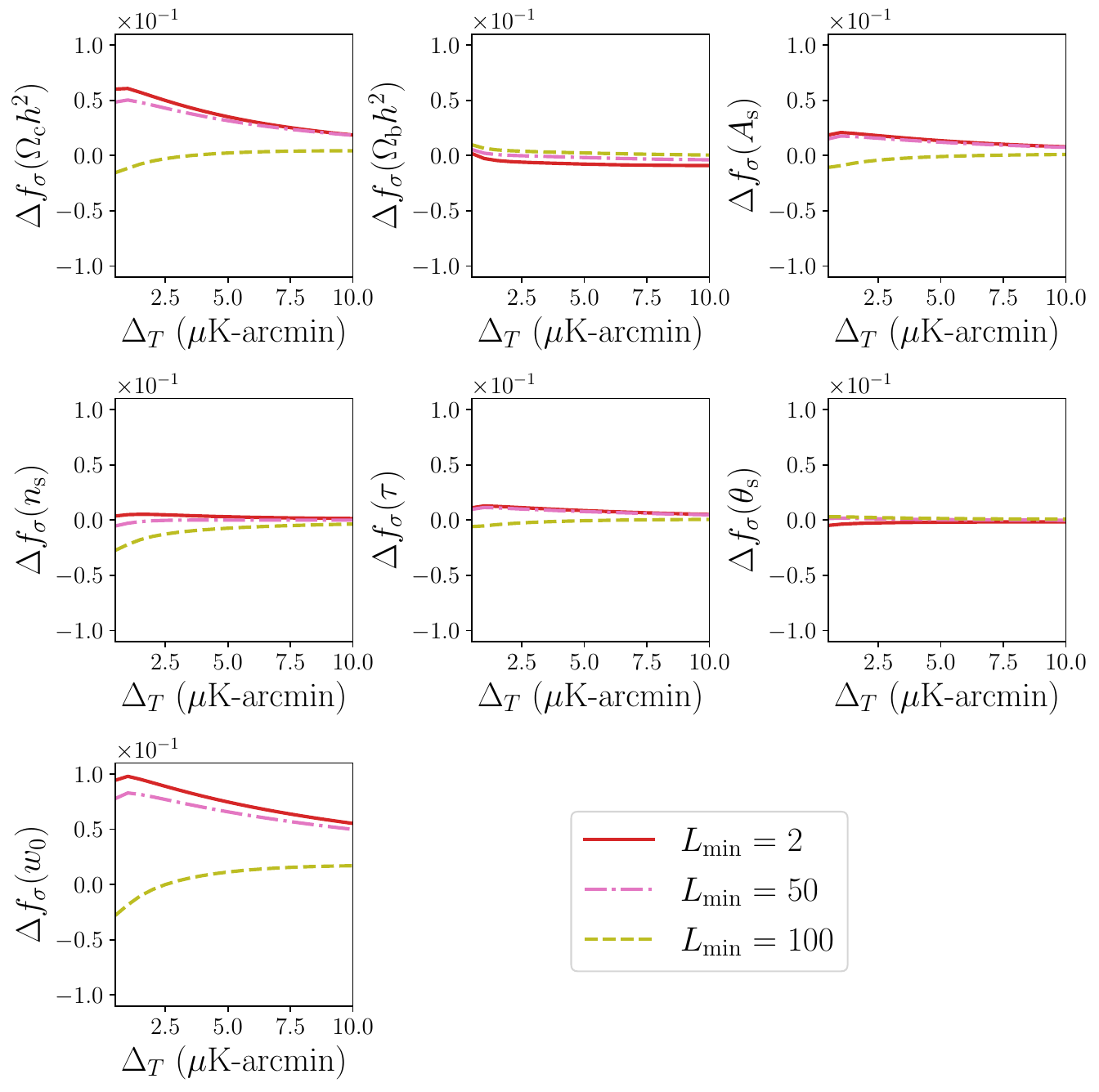}
    \caption{The fractional difference in the error bar when the covariances between the primary CMB and the reconstructed lensing potential are neglected (defined in eq.~\ref{eq:FractionalDiff}), for a 7-parameter $\Lambda$CDM + $w_0$ model. For most parameters, the shift is less than 3\%. For $w_0$, the difference can be as large as 5\% even at 10$\mu$K-arcmin noise levels, if we include modes of $L = 50$ or below. The sign of the change can be either positive or negative due to the complicated structure of the correlations in fig.~\ref{fig:Correlations}, which affect the Fisher information in a nontrivial way.}
    \label{fig:ParamDiffsLensed_w}
\end{figure}

In figure~\ref{fig:ParamConstraintsLensed_w}, we present results from a 7-parameter forecast that includes $w_0$, for $L_\mathrm{min} = 2$ in the lensing power spectrum. The dashed green line shows results when including Gaussian covariances only, the dashed-dotted purple line includes lensing-induced non-Gaussian covariances amongst primary CMB modes, and the solid orange line additionally includes lensing-induced non-Gaussian covariances between the primary CMB and the reconstructed lensing power. For most parameters, although the total non-Gaussian covariances can have a nontrivial impact on the parameter constraints, the inclusion of ${\rm Cov}_{\ell_1\ell_2}^{IJ,\phi\phi}$ has little effect in comparison; for $w_0$, however, this term becomes important, and to a lesser extent for $\Omega_\mathrm{c} h^2$ as well. In figure~\ref{fig:ParamDiffsLensed_w}, we show the fractional difference in the error bar,
\begin{equation}
    \Delta f_\sigma = \frac{\sigma(\mathrm{CMBxCMB + CMBxL}) - \sigma(\mathrm{CMBxCMB})}{\sigma(\mathrm{CMBxCMB + CMBxL})} ,
    \label{eq:FractionalDiff}
\end{equation}
where $\sigma(\mathrm{CMBxCMB + CMBxL})$ is the 1-$\sigma$ error bar when all non-Gaussian covariances are included, and $\sigma(\mathrm{CMBxCMB})$ is the 1-$\sigma$ error bar when the only non-Gaussian covariances that are included are those amongst primary CMB $\ell$-modes. We present results for three different values of $L_\mathrm{min}$: 2, 50, and 100. In general, the importance of ${\rm Cov}_{\ell_1\ell_2}^{IJ,\phi\phi}$ tends to increase slightly at lower noise levels for a given $L_\mathrm{min}$; for $w_0$, its impact is already on the order of 5\% of the error bar at a noise level of 10.0 $\mu$K-arcmin when including $L_\mathrm{min} = 50$, and increases by a few percent as the noise decreases. 

We note that although dropping the correlations between the lensed CMB and lensing power removes some correlations from the full covariance matrix, there are some cases where the error bar worsens, resulting in a negative fractional change. As seen in fig.~\ref{fig:Correlations}, the correlation matrix between the lensed CMB and lensing power has a complex structure, including both positive and negative terms. Including or neglecting these terms changes the Fisher information in a non-trivial way, and therefore affects the parameter error bars in a complicated manner that may cause them to either increase or decrease.

\begin{figure}[t!]
    \centering
    \includegraphics[width=0.95\columnwidth]{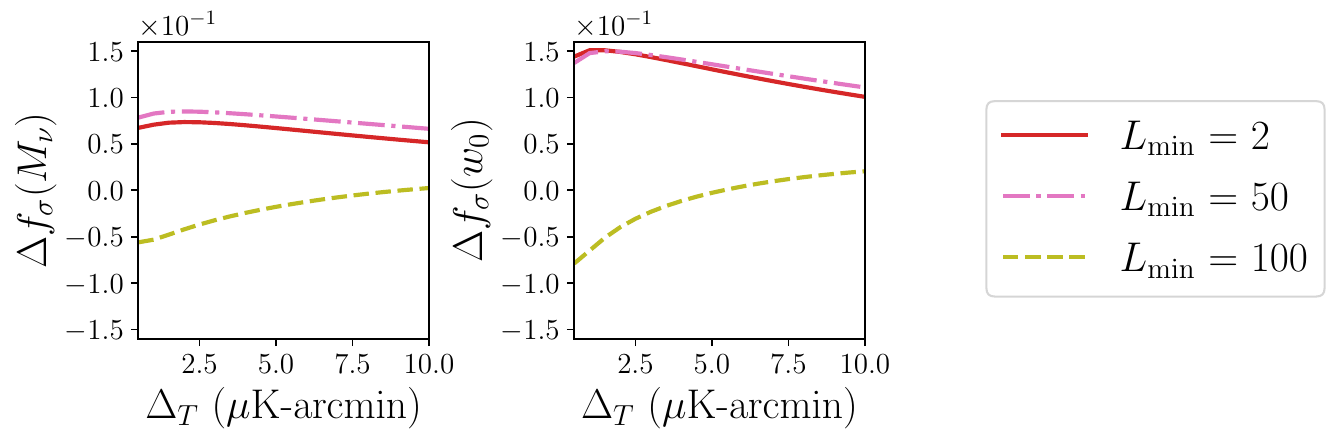}
    \caption{The fractional difference in the error bar when the covariances between the primary CMB and the reconstructed lensing potential are neglected (defined in eq.~\ref{eq:FractionalDiff}), for an 8-parameter $\Lambda$CDM + $\Sigma m_{\nu}$ + $w_0$ model.}
    \label{fig:ParamDiffsLensed_mw}
\end{figure}

\begin{figure}[t!]
    \centering
    \includegraphics[width=0.95\columnwidth]{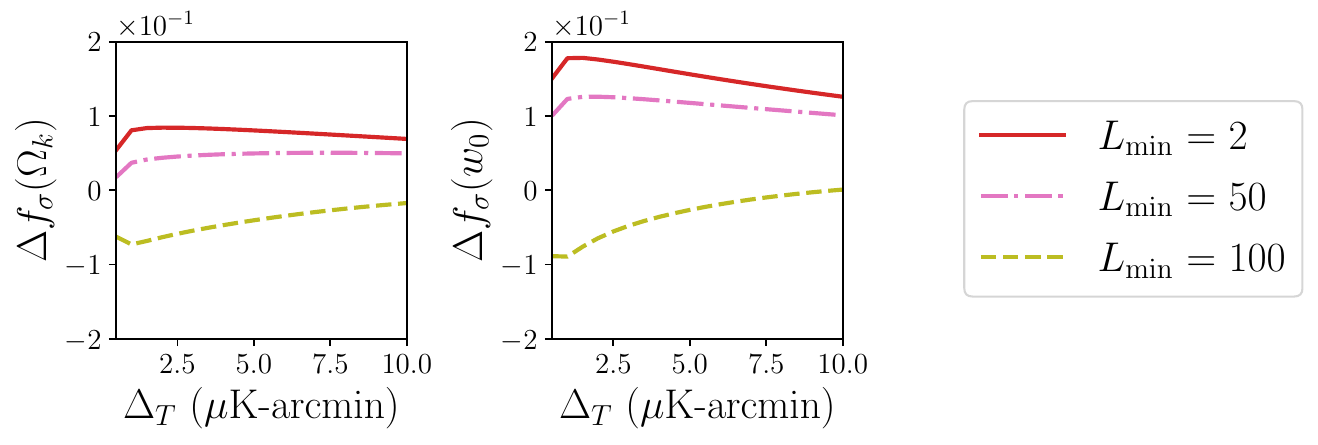}
    \caption{The fractional difference in the error bar when the covariances between the primary CMB and the reconstructed lensing potential are neglected (defined in eq.~\ref{eq:FractionalDiff}), for an 8-parameter $\Lambda$CDM + $\Omega_\mathrm{k}$ + $w_0$ model.}
    \label{fig:ParamDiffsLensed_kw}
\end{figure}

If we take the $w_0$ model and furthermore add additional parameters which are significantly informed by lensing information, the impact of the lensed CMB and lensing power correlations becomes even more important. In figure~\ref{fig:ParamDiffsLensed_mw}, we show results for the fractional difference, eq.~\ref{eq:FractionalDiff}, for a 2-parameter extension of $\Lambda$CDM + $\Sigma m_{\nu}$ + $w_0$, while in figure~\ref{fig:ParamDiffsLensed_kw} we present $\Lambda$CDM + $\Omega_\mathrm{k}$ + $w_0$. In all cases, there is some weak dependence on noise, but more importantly, the significance of the correlations between the lensed CMB and the reconstructed lensing power is captured to differing degrees depending on which lensing $L$-modes are included in the forecast. For $L_\mathrm{min} = 100$, the effect of the ${\rm Cov}_{\ell_1\ell_2}^{IJ,\phi\phi}$ terms is relatively small, except at particularly low noise levels around 1$\mu$K-arcmin. However, for $w_0$, these terms will matter noticeably at all noise levels when the analysis includes $L_\mathrm{min} = 50$, similar to what is seen in figure~\ref{fig:ParamDiffsLensed_w}. When varying both $w_0$ and $\Omega_k$, neglecting the ${\rm Cov}_{\ell_1\ell_2}^{IJ,\phi\phi}$ terms can result in mis-estimating the error bar on $w_0$ by almost 20\% if we include $L_\mathrm{min} = 2$. 

\begin{figure}[t!]
    \centering
    \includegraphics[width=0.95\columnwidth]{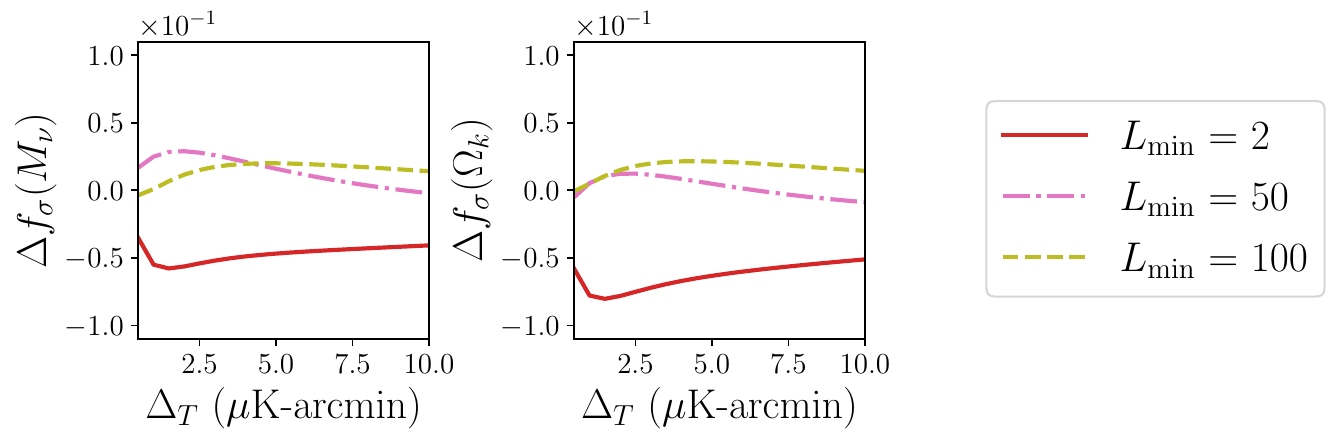}
    \caption{The fractional difference in the error bar when the covariances between the primary CMB and the reconstructed lensing potential are neglected (defined in eq.~\ref{eq:FractionalDiff}), for an 8-parameter $\Lambda$CDM + $\Sigma m_{\nu}$ + $\Omega_\mathrm{k}$ model.}
    \label{fig:ParamDiffsLensed_mk}
\end{figure}

When adding $\Sigma m_{\nu}$ or $\Omega_\mathrm{k}$ one at a time as single-parameter extensions to $\Lambda$CDM, we find that the effect of the ${\rm Cov}_{\ell_1\ell_2}^{IJ,\phi\phi}$ terms on parameter constraints is small, at most on the order of 3\% of the parameter error bars. If we include the pair of these parameters as an extension, however, the ${\rm Cov}_{\ell_1\ell_2}^{IJ,\phi\phi}$ terms can become important, due to the degeneracies between parameters and the information carried by the low $L$ modes. For the 2-parameter extension of $\Lambda$CDM + $\Sigma m_{\nu}$ + $\Omega_\mathrm{k}$, we plot the fractional difference in the error bars in figure~\ref{fig:ParamDiffsLensed_mk}. We find that the correlations of ${\rm Cov}_{\ell_1\ell_2}^{IJ,\phi\phi}$ can matter on the order of 5\% or more. Unlike the models where we included the dark energy equation of state, we find that for the combination of neutrino mass and curvature, the most significant contributions from ${\rm Cov}_{\ell_1\ell_2}^{IJ,\phi\phi}$ are relevant only for the lowest $L$-modes and have less of an impact when $L_\mathrm{min} = 50$ compared to the extensions that included $w_0$.

\section{Conclusion}
\label{sec:Conclusion}

Gravitational lensing of the Cosmic Microwave Background generates off-diagonal covariances between $\ell$-modes of the primary CMB power spectra, and cross-covariances between the primary CMB and the lensing power that is reconstructed from the CMB. Neglecting these non-Gaussian covariances in their entirety can result in parameter constraints that are too optimistic on the order of 10\% or more, and modeling them will thus be important for future CMB data analyses. In this work, we have used Fisher forecasting to test, for various experimental configurations relevant for both current and future CMB experiments, the impact of the cross-covariances between primary CMB and reconstructed lensing power spectra.

We find that the full model of the non-Gaussian covariances, including off-diagonal $\ell \times \ell$ correlations amongst primary CMB modes, can be important at the level of 10\% for getting accurate parameter constraints on $\Lambda$CDM, consistent with prior work studying these terms. Compared to this total effect, we find that the cross-covariances of the primary CMB with the lensing power spectrum, specifically, are a small contribution to error bars on $\Lambda$CDM parameters; neglecting these terms in our forecasts impacts parameter error bars on the order of 1\%, and by no more than 3\%, for a range of noise levels from 0.5 to 10.0 $\mu$K-arcmin in the temperature, with a 1.4 arcmin beam. For a $\Lambda$CDM model, accounting for the $\ell \times \ell$ correlations of the primary CMB will be the more important factor in accurately determining parameter constraints from lensed data.

For parameter extensions beyond $\Lambda$CDM, the significance of the ${\rm Cov}_{\ell_1\ell_2}^{IJ,\phi\phi}$ covariances will depend on the parameters, their degeneracies, and to what extent different lensing $L$-modes are contributing to the estimation of the parameters. We find that for the dark energy equation of state $w_0$, in particular, even a single-parameter extension to $\Lambda$CDM results in a parameter space where the correlations between the lensed CMB and reconstructed lensing power have a nontrivial impact on parameter constraints, as much as 10\% for $w_0$ at low noise levels and with the inclusion of lensing modes down to $L = 2$. Further extending the 7-parameter $w_0$ model with other parameters informed by lensing can make these correlations even more significant, as much as 20\% of the error bar at low noise and low $L$-modes. Without including the dark energy equation of state as a free parameter, adding both neutrino mass and curvature to $\Lambda$CDM also results in a model where the ${\rm Cov}_{\ell_1\ell_2}^{IJ,\phi\phi}$ covariances can matter on the order of 5\% or more of the error bar, when $L_\mathrm{min} = 2$. CMB surveys that are doing analyses with these free parameters should take care of the size of this effect; it could be accounted for by including the correlations of equation~\ref{eq:covarianceXXdd}, between the lensed CMB and the reconstructed lensing power, directly in the likelihood, or mitigated in other ways, such as delensing the CMB data in order to reduce the impact of the lensing-induced non-Gaussian covariances as a whole.

\section*{Acknowledgments}
We thank Lennart Balkenhol, Silvia Galli, Gil Holder, Joel Meyers, Srinivasan Raghunathan, and the South Pole Telescope (SPT) collaboration for helpful discussions and feedback. This work was supported by the Center for AstroPhysical Surveys (CAPS) at the National Center for Supercomputing Applications (NCSA), University of Illinois Urbana-Champaign. This work made use of the Illinois Campus Cluster, a computing resource that is operated by the Illinois Campus Cluster Program (ICCP) in conjunction with the National Center for Supercomputing Applications (NCSA) and which is supported by funds from the University of Illinois at Urbana-Champaign.

\bibliographystyle{utphys}
\bibliography{correlations}

\end{document}